%% file: main.tex
\documentclass[10pt,conference,screen=true]{IEEEtran}
\IEEEoverridecommandlockouts

\usepackage{cite}
\usepackage{graphicx}
\usepackage{textcomp}
\usepackage[dvipsnames]{xcolor}
\usepackage{comment}
\usepackage{multirow}
\usepackage{booktabs}
\usepackage{threeparttable}
\usepackage{pgfplots}
\pgfplotsset{compat=newest}
\usepackage{subfig}
\usepackage{subcaption}

\usepackage{physunits}
\usepackage{siunitx}
\graphicspath{{./figs/}{../}}

\makeatletter
\newif\if@restonecol
\makeatother
\usepackage{algorithm}
\usepackage{algpseudocode}
\usepackage{amsmath,amssymb,amsfonts}

\usepackage{cleveref}
\Crefformat{figure}{Fig.~#2#1#3}
\Crefname{subfigure}{Fig.}{Figs.}
\Crefname{figure}{Fig.}{Figs.}
\Crefname{section}{Section}{Sections}
\usepackage{url}
\usepackage{tikz}
\usepackage{enumitem}

\newcommand{\minisection}[1]{\vspace{.02in}\noindent{\textbf{#1}}.}

\newcommand{\vr}[2]{#1\,{\scriptsize(#2$\times$)}}

\usepackage{amsthm}

\newtheoremstyle{defnobold}{}{}{}{}{\itshape}{.}{.5em}{}%
\theoremstyle{defnobold}

\def\BibTeX{{\rm B\kern-.05em{\sc i\kern-.025em b}\kern-.08em
    T\kern-.1667em\lower.7ex\hbox{E}\kern-.125emX}}

\begin{document}

\title{
  Chiplet3D: Pin- and Thermal-Aware 3D Chiplet Floorplanning via Convolution-Embedded MILP
}

\author{
  \IEEEauthorblockN{Shuo Ren, Libo Shen, Yaohui Han, Rongliang Fu, Junying Huang, Bei Yu and Tsung-Yi Ho}
}

\maketitle
\thispagestyle{plain}
\pagestyle{plain}

\begin{abstract}
  \input{doc/0-abstract}
\end{abstract}

\input{doc/1-intro}
\input{doc/2-prelim}
\input{doc/3-method}
\input{doc/4-exp}

\input{doc/5-conclusion}

\bibliographystyle{IEEEtran}
\bibliography{ref/Top-sim, ref/reference}

\end{document}

%% file: doc/0-abstract.tex
As traditional Moore's Law scaling slows down, 3D-ICs stack multiple active dies vertically to sustain performance scaling. However, this vertical stacking traps heat inside, making temperature a design concern. Although we can fix thermal issues at different design steps, floorplanning is the earliest and most cost-effective stage to solve it. Previous methods handle this by assuming wires connect to block centers and estimating temperature through simplistic power-based calculations, but these assumptions mislead their wirelength optimization and leave hotspots unresolved.
To address these limitations, we present Chiplet3D, a pin- and thermal-aware floorplanner for two-die 3D-ICs. To achieve pin-awareness, it supports all four rotations and two flips, measuring wirelength from exact pin locations so the solver can flip or rotate blocks to pull connected pins closer. On the thermal side, Chiplet3D replaces the inaccurate power-based metrics of prior work with a fast, coarse convolution field embedded directly in a mixed-integer linear program (MILP) to accurately track the true 3D heat spread. 
We evaluate Chiplet3D on the ICCAD'24 ATPlace benchmarks, validating every temperature with a golden 3D-ICE simulation. 
Chiplet3D reduces wirelength by 39\%--43\% on average (and up to 62\% in the best case), while lowering peak temperatures by up to 45.9$^\circ$C and reducing thermal non-uniformity by up to 56\% compared to the SOTA baselines. Overall, these results demonstrate that by co-optimizing pin alignment and thermal fields, Chiplet3D establishes a stronger Pareto frontier between thermal-aware layout and interconnect efficiency.

%% file: doc/1-intro.tex
\section{Introduction}
\label{sec:intro}
\begin{figure}[t!]
      \centering
      \includegraphics[width=0.95\linewidth]{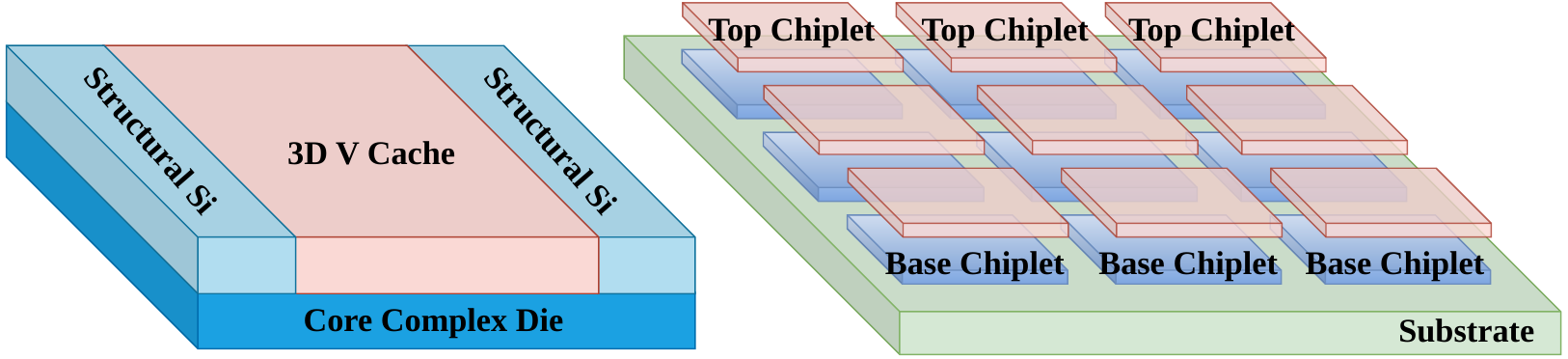}%
      \caption{Left: AMD's 3D V-Cache~\cite{amd3dvcache} illustrates a commercial stacked-die product. Right: 3D chiplet floorplanning model.}
      \label{fig:intro}
\end{figure}

As semiconductor technology approaches the physical limits of planar scaling~\cite{wong2025quest, neisser2021international}, three-dimensional integrated circuits (3D-ICs) have emerged as a key way to keep performance scaling~\cite{debenedictis2017sustaining}.
By vertically integrating functional dies with 3D interconnect technologies~\cite{xia2025tsv, nam2025hybrid}, 3D-ICs shorten global interconnects and raise integration density~\cite{chen2025survey25d, pasricha2009exploring}.
The approach is no longer experimental: commercial parts such as AMD's 3D V-Cache~\cite{amd3dvcache} (\Cref{fig:intro}, left), TSMC's SoIC~\cite{chen2024soic}, and Intel's Foveros~\cite{ingerly2019foveros}, together with interconnect standards like UCIe~\cite{sharma2022ucie}, have made vertical stacking a mainstream option for high-performance systems~\cite{chao2024ai3dic, hachemi2025advanced, lee2021soic, lau2024recent}.

This increase in integration density, however, traps heat: stacked dies leave few escape paths~\cite{wei2024tim}, so heat builds up in the interior layers~\cite{koroglu2023thermal, wei2024tim, salvi2021thermal, guo2025thermal}, and heterogeneous stacks that press a high-power compute die against memory or I/O tiles develop steep temperature gradients.
The resulting hotspots raise leakage~\cite{liu2007leakage, sultan2018leakage}, degrade timing~\cite{zhan2008thermally, ajami2005modeling} and reliability~\cite{mishra2016em}, and force chips to throttle below their target frequency~\cite{liu2007leakage, zhan2008thermally, brooks2001dynamic}.

Therefore, 3D floorplanning represents the earliest and most cost-effective stage in 3D-IC design to tackle these thermal issues. This single stage simultaneously determines die-to-die vertical overlap, lateral spacing between hot blocks, and inter-layer power balance, all while safeguarding high system performance by minimizing interconnect length~\cite{cong2006thermal, cuesta2015thermal}. Consequently, a good 3D floorplanner must co-optimize both wirelength and temperature.
Researchers have made extensive efforts in thermal‑aware floorplanning for 3D‑ICs.
Early work focused on monolithic integration flows for stacking existing 2D IP blocks~\cite{2013ASPDAC_Lim_3dflp}. Subsequent fixed- and free-outline methods have increasingly co-optimized interconnect cost with layer assignment, TSV planning, and thermal objectives~\cite{cong2006thermal, cuesta2015thermal, 2021TVLSI_t3dflp, 2023TVLSI_t3dflp}. Recent flows have extended these goals to heterogeneous chiplet integration and multi-objective floorplanning~\cite{chen2023floorplet, 2024ICCAD-ATPlace2.5D, ren2026partitioning, zhuang2022multi, zhuang2023multi}.
Lin et al.~\cite{2021TVLSI_t3dflp} proposed an analytical fixed-outline floorplanner that jointly optimizes wirelength and temperature by adding thermal forces into a continuous placement formulation, while
Guan et al.~\cite{2023TVLSI_t3dflp} improved thermal-aware 3-D floorplanning by using power-density-aware tier assignment and analytical placement to obtain a better cross-tier power distribution.

However, these prior methods still leave three limitations during optimization.
(1) \textbf{Center-based connectivity modeling.} Wirelength is estimated from block centers rather than true pin locations, so the optimizer cannot see how rotations and flips move connected pins closer together.
(2) \textbf{Restricted block transformations.} To keep the search tractable, many methods restrict or drop rotation and flip choices, discarding orientations that could improve both pin-to-pin wirelength and thermal distribution.
(3) \textbf{Simplified thermal estimation.} Thermal objectives are based on coarse in-optimizer temperature estimates, while accurate full-stack thermal behavior is only checked after placement; consequently, hotspot geometry is only partially resolved during search.

To address these limitations, Chiplet3D introduces a pin- and thermal-aware 3-D chiplet floorplanner.
It models HPWL from transformed pin coordinates under rotations and flips, allowing the solver to directly optimize pin-to-pin connectivity, and embeds a compact convolution-based thermal model into the MILP so that thermal spreading is considered during optimization efficiently rather than only after placement.
We evaluate Chiplet3D on the ICCAD'24 ATPlace~\cite{2024ICCAD-ATPlace2.5D} benchmarks and validate the temperature of every layout with a golden 3D-ICE simulation~\cite{zhu20253d}.
Overall, our main contributions are summarized as follows:
\begin{itemize}[leftmargin=1.2em, itemsep=0.2em]
    \item \textbf{Pin-aware modeling.} We propose a pin-aware wirelength model that makes the HPWL over true pin locations directly solvable in the MILP, so the solver itself decides how rotations and flips pull connected pins closer (see~\Cref{fig:hpwl_comparison}).
    \item \textbf{Coarse convolutional thermal field.} We embed a coarse convolutional model of heat conduction in the MILP, letting the solver do thermal-aware floorplanning without calling a thermal simulator, while locating the golden 3D-ICE hotspot to within one grid cell (\Cref{tab:fidelity}).
    \item \textbf{Dynamic thermal elasticity.} Chiplet3D fixes a cool thermal skeleton, then spends just enough floorplan freedom on wirelength optimization to shorten wires without surrendering the skeleton's thermal advantage.
    \item \textbf{MILP-compatible co-optimization with strong empirical gains.} Chiplet3D shortens wirelength by 39\%--43\% on average (up to 62\%) and cuts the verified peak temperature by up to 45.9$^\circ$C and thermal non-uniformity by up to 56\%, while winning 8--9 of the 10 cases on bottom-die peak and thermal uniformity.
\end{itemize}

%% file: doc/2-prelim.tex
\section{Preliminaries}
\label{sec:prelim}
\subsection{3D Chiplet Floorplanning}
\label{sec:problem_def}
Two-die 3D chiplet floorplanning arranges a set of blocks across two vertically stacked dies, and the resulting layout is judged by how short its wiring is and how cool it runs, the two demands raised in \Cref{sec:intro}.
Formally, a floorplanning instance is a triple $\mathcal{F}=(B,N,L)$ comprising a block set, a netlist, and a layer set.
The block set $B=\{b_1,b_2,\ldots,b_n\}$ contains rectangular blocks; each block $b_i$ has input width $w_i$, height $h_i$, and power consumption $p_i$. The task is to decide, for every block, where it sits and whether it lands on the top or the bottom die.

The netlist $N=\{n_1,n_2,\ldots,n_m\}$ wires these blocks together, where each net $n_j$ connects a subset of blocks $B_j\subseteq B$.
The layer set $L=\{l_0,l_1\}$ represents the two active dies, which share a common outline $W\times H$.

A solution assigns each block $b_i$ a bottom-left position $(x_i,y_i)$, a die assignment $z_i\in\{0,1\}$ (block $b_i$ lands on layer $l_{z_i}\in L$), an orientation $o_i\in\{\mathrm{R0},\mathrm{R90},\mathrm{R180},\mathrm{R270}\}$, and horizontal/vertical flips $f_i=(f_i^h,f_i^v)\in\{0,1\}^2$.
The orientation $o_i$ sets a block's effective footprint, which the placement constraints depend on.
Writing the rotation angle as $\theta_o\in\{0,\pi/2,\pi,3\pi/2\}$ for $o\in\{\mathrm{R0},\mathrm{R90},\mathrm{R180},\mathrm{R270}\}$, the effective dimensions of block $b_i$ after applying orientation $o_i$ are shown in \Cref{eq:eff_dims}.
\begin{equation}
  \begin{aligned}
    w'_i &= w_i\,|\cos\theta_{o_i}| + h_i\,|\sin\theta_{o_i}|, \\
    h'_i &= h_i\,|\cos\theta_{o_i}| + w_i\,|\sin\theta_{o_i}|. 
  \end{aligned}
  \label{eq:eff_dims}
\end{equation}
A $90^\circ$ or $270^\circ$ rotation thus swaps a block's width and height, whereas $0^\circ$ and $180^\circ$ leave them unchanged; the floorplan constraints act on these effective dimensions $(w'_i,h'_i)$, not the nominal $(w_i,h_i)$ when rotations are taken into consideration.

\subsection{Pin-Aware Wirelength}
\label{sec:pin_prelim}
\input{figs/hpwl_comparison}
\Cref{fig:hpwl_comparison} motivates the need for a pin-accurate wirelength model. Building each net's bounding box from transformed pin coordinates, rather than from block centers, changes both the wirelength and the ranking of different orientations. A center-based proxy ignores intra-block pin geometry and can misrank configurations. This effect is especially pronounced for blocks with many or asymmetrically placed pins, where orientation becomes a real lever for shortening nets that a center-based objective cannot capture.
Blocks connect through pins at fixed offsets from the block center, not through the centers themselves.
We write a pin as a pair $(i,k)\in n_j$ for the $k$-th pin of block $b_i$, and let $(u_{i,k}, v_{i,k})$ be its input offset from the block center.
Rotating and then flipping the block maps this offset to
\begin{align}
  (u'_{i,k},\, v'_{i,k}) = T_f\big(T_r(u_{i,k}, v_{i,k}, o_i),\, f_i^h, f_i^v\big), \label{eq:pin_chain}
\end{align}
where the flip is a sign change, $T_f(x, y, f^h, f^v) = ((-1)^{f^h} x,\, (-1)^{f^v} y)$, and the rotation is
\begin{align}
  T_r(x,y,o) = (x\cos\theta_o-y\sin\theta_o,\; x\sin\theta_o+y\cos\theta_o). \label{eq:rot_transform}
\end{align}
Here $\theta_o\in\{0,\tfrac{\pi}{2},\pi,\tfrac{3\pi}{2}\}$ is the rotation angle of orientation $o_i$ (\Cref{eq:eff_dims}) and $f_i^h,f_i^v\in\{0,1\}$ are the horizontal and vertical flips, so \Cref{eq:pin_chain} first rotates a local pin offset and then mirrors it; the result $(u'_{i,k},v'_{i,k})$ is where that pin lands once block $b_i$ is reoriented.

Each block has its own pin pattern, which is often asymmetric. The pin offset, and hence the net bounding box, depends on the chosen transform, precisely the dependence a center-based estimate discards.
The HPWL of a net is the half-perimeter of the bounding box enclosing its terminals. The conventional center‑based model builds this bounding box from the connected blocks' centers.
Let $(x_i^c,y_i^c)=(x_i+w'_i/2,\,y_i+h'_i/2)$ be the center of block $b_i$ after orientation.
For net $n_j$, define its center-based spans as
\begin{equation}
    \begin{aligned}
  \Delta x_j^{\mathrm{ctr}} &= \max_{i\in B_j} x_i^c - \min_{i\in B_j} x_i^c, \\
  \Delta y_j^{\mathrm{ctr}} &= \max_{i\in B_j} y_i^c - \min_{i\in B_j} y_i^c. 
    \end{aligned}
    \label{eq:ctr_span}
\end{equation}
The center-based estimate is then
\begin{align}
  \mathrm{HPWL}_{\mathrm{ctr}} = \sum_{n_j\in N}\left(\Delta x_j^{\mathrm{ctr}}+\Delta y_j^{\mathrm{ctr}}\right).
  \label{eq:hpwl_center}
\end{align}
\input{tables/table_notation}
This proxy is invariant to in-place rotations that leave the block center unchanged.
Pin locations, however, move under rotations and flips, and these movements directly change net bounding boxes and pairwise Manhattan distances.
For pin $k$ on block $b_i$, the transformed pin offset of \Cref{eq:pin_chain} gives the global pin position
\begin{align}
  X_{i,k} &= x_i^c + u'_{i,k}, &
  Y_{i,k} &= y_i^c + v'_{i,k}. \label{eq:pin_global_prelim}
\end{align}
For the same net, the pin-based spans are
\begin{equation}
\begin{aligned}
  \Delta x_j^{\mathrm{pin}} &= \max_{(i,k)\in n_j} X_{i,k} - \min_{(i,k)\in n_j} X_{i,k}, \\
  \Delta y_j^{\mathrm{pin}} &= \max_{(i,k)\in n_j} Y_{i,k} - \min_{(i,k)\in n_j} Y_{i,k}. 
\end{aligned}
\label{eq:pin_span}
\end{equation}
The pin-aware HPWL used in this work is therefore
\begin{align}
  \mathrm{HPWL}_{\mathrm{pin}} = \sum_{n_j\in N}\left(\Delta x_j^{\mathrm{pin}}+\Delta y_j^{\mathrm{pin}}\right).
  \label{eq:hpwl_pin}
\end{align}
The pin-aware $\mathrm{HPWL}$ is the wirelength model used throughout this work and reported in~\Cref{sec:exp_setup}.
\subsection{Multi-Objective Optimization Formulation}
\label{sec:multi_objective}
\noindent The symbols introduced above are summarized in \Cref{tab:notation}. With them, we state the thermal-aware 3D chiplet floorplanning problem by its input, output, and objective.

\minisection{Input}
A floorplanning instance $\mathcal{F}=(B,N,L)$ of \Cref{sec:problem_def}: a block set $B$ in which every $b_i$ carries a width $w_i$, height $h_i$, and power $p_i$; a netlist $N$; and two vertically stacked dies $L=\{l_0,l_1\}$ sharing a common $W\times H$ outline.

\minisection{Output}
An optimized 3D floorplan layout of each block: its position $(x_i,y_i)$, die $z_i$, orientation $o_i$, and flips $f_i^h,f_i^v$, that stays inside the outline,
\begin{align}
  0 \le x_i,\quad x_i + w'_i \le W, \quad 0 \le y_i, \quad y_i + h'_i \le H, \label{eq:outline}
\end{align}
and does not overlap with any other block on the same die,
\begin{equation}
\begin{aligned}
  &(x_i + w'_i \le x_j) \;\lor\; (x_j + w'_j \le x_i) \;\lor\; \\
  &\qquad (y_i + h'_i \le y_j) \;\lor\; (y_j + h'_j \le y_i),
\end{aligned}
\label{eq:nonoverlap}
\end{equation}
where the effective dimensions $(w'_i,h'_i)$ track the orientation through \Cref{eq:eff_dims}; orientation-uniqueness and configuration-feasibility constraints complete the feasible set.

\minisection{Objective}
Find a floorplan that minimizes the weighted sum of wirelength and peak temperature,
\begin{align}
  \min \quad & \alpha\, \mathrm{HPWL}\; +\; (1-\alpha)\, t_{\mathrm{peak}}, \label{eq:objective}
\end{align}
where $\mathrm{HPWL}$ is the pin-aware half-perimeter wirelength defined in \Cref{eq:hpwl_pin} and $t_{\mathrm{peak}}$ is the peak on-die temperature. A single weight $\alpha\in[0,1]$ trades off the two objectives: a good 3D chiplet floorplan must keep wires short and heat under control at the same time.

%% file: figs/hpwl_comparison.tex
\usetikzlibrary{backgrounds}
\newcommand{\figrightshift}{0pt}
\begin{figure}[t!]
  \centering
  \hspace*{\figrightshift}%
  \resizebox{0.95\linewidth}{!}{%
  \begin{tabular}{@{}c@{\hskip 0.27cm}c@{}}
  \begin{tikzpicture}[scale=0.545,
    block/.style={draw, thick, fill=gray!15},
    bbox/.style={dashed, thick, Maroon},
    dim/.style={<->, >=stealth, semithick, RoyalBlue},
    lbl/.style={font=\footnotesize},
    rowlbl/.style={font=\footnotesize\itshape, text=gray!70!black},
    midlbl/.style={font=\footnotesize},
    show background rectangle,
    inner frame sep=2pt,
    background rectangle/.style={fill=blue!9, draw=none, rounded corners=3pt},
  ]
    \def\ty{0}
    \draw[block] (0,\ty) rectangle (2.2,\ty+1.2);
    \node[lbl] at (1.1,\ty+0.9) {$b_1$};
    \draw[block] (3.2,\ty+1.4) rectangle (4.6,\ty+3.6);
    \node[lbl] at (3.9,\ty+3.15) {$b_2$};
    \foreach \cx/\cy in {1.1/0.6, 3.9/2.5} {
      \draw[thick] (\cx-0.11,\ty+\cy-0.11) -- (\cx+0.11,\ty+\cy+0.11);
      \draw[thick] (\cx-0.11,\ty+\cy+0.11) -- (\cx+0.11,\ty+\cy-0.11);
    }
    \draw[bbox] (1.1,\ty+0.6) rectangle (3.9,\ty+2.5);
    \draw[dim] (1.1,\ty-0.4) -- node[below,lbl] {$\Delta x$} (3.9,\ty-0.4);
    \draw[dim] (5.2,\ty+0.6) -- node[right,lbl] {$\Delta y$} (5.2,\ty+2.5);
    \draw[dotted,gray] (1.1,\ty+0.6) -- (1.1,\ty-0.4);
    \draw[dotted,gray] (3.9,\ty+0.6) -- (3.9,\ty-0.4);
    \draw[dotted,gray] (4.6,\ty+0.6) -- (5.2,\ty+0.6);
    \draw[dotted,gray] (4.6,\ty+2.5) -- (5.2,\ty+2.5);
    \node[rowlbl, anchor=east] at (1.0,\ty+1.8) {original};
    \node[midlbl] at (2.5,-1.5) {HPWL stays the same};
    \def\by{-5.2}
    \draw[block] (0,\by) rectangle (2.2,\by+1.2);
    \node[lbl] at (1.1,\by+0.9) {$b_1$};
    \draw[block] (2.8,\by+1.8) rectangle (5.0,\by+3.2);
    \node[lbl] at (3.9,\by+2.85) {$b_2'$};
    \foreach \cx/\cy in {1.1/0.6, 3.9/2.5} {
      \draw[thick] (\cx-0.11,\by+\cy-0.11) -- (\cx+0.11,\by+\cy+0.11);
      \draw[thick] (\cx-0.11,\by+\cy+0.11) -- (\cx+0.11,\by+\cy-0.11);
    }
    \draw[bbox] (1.1,\by+0.6) rectangle (3.9,\by+2.5);
    \draw[dim] (1.1,\by-0.4) -- node[below,lbl] {$\Delta x$} (3.9,\by-0.4);
    \draw[dim] (5.2,\by+0.6) -- node[right,lbl] {$\Delta y$} (5.2,\by+2.5);
    \draw[dotted,gray] (1.1,\by+0.6) -- (1.1,\by-0.4);
    \draw[dotted,gray] (3.9,\by+0.6) -- (3.9,\by-0.4);
    \draw[dotted,gray] (5.0,\by+0.6) -- (5.2,\by+0.6);
    \draw[dotted,gray] (5.0,\by+2.5) -- (5.2,\by+2.5);
    \node[rowlbl, anchor=east] at (1.0,\by+1.8) {rotation};
    \node[font=\small] at (2.5,\by-1.25) {(a) Block-center HPWL};
  \end{tikzpicture}
  &
  \begin{tikzpicture}[scale=0.545,
    block/.style={draw, thick, fill=gray!15},
    pin/.style={circle, fill=black, inner sep=1.3pt},
    bbox/.style={dashed, thick, Maroon},
    dim/.style={<->, >=stealth, semithick, RoyalBlue},
    lbl/.style={font=\footnotesize},
    rowlbl/.style={font=\footnotesize\itshape, text=gray!70!black},
    midlbl/.style={font=\footnotesize},
    show background rectangle,
    inner frame sep=2pt,
    background rectangle/.style={fill=orange!13, draw=none, rounded corners=3pt},
  ]
    \def\ty{0}
    \draw[block] (0,\ty) rectangle (2.2,\ty+1.2);
    \node[lbl] at (1.1,\ty+0.9) {$b_1$};
    \draw[block] (3.2,\ty+1.4) rectangle (4.6,\ty+3.6);
    \node[lbl] at (3.9,\ty+3.15) {$b_2$};
    \node[pin] (p1t) at (1.85,\ty+0.3) {};
    \node[pin] (p2t) at (4.3,\ty+3.3) {};
    \draw[bbox] (1.85,\ty+0.3) rectangle (4.3,\ty+3.3);
    \draw[dim] (1.85,\ty-0.4) -- node[below,lbl] {$\Delta x$} (4.3,\ty-0.4);
    \draw[dim] (5.2,\ty+0.3) -- node[right,lbl] {$\Delta y$} (5.2,\ty+3.3);
    \draw[dotted,gray] (1.85,\ty+0.3) -- (1.85,\ty-0.4);
    \draw[dotted,gray] (4.3,\ty+0.3) -- (4.3,\ty-0.4);
    \draw[dotted,gray] (4.6,\ty+0.3) -- (5.2,\ty+0.3);
    \draw[dotted,gray] (4.6,\ty+3.3) -- (5.2,\ty+3.3);
    \node[rowlbl, anchor=east] at (1.0,\ty+1.8) {original};
    \node[midlbl] at (2.5,-1.5) {HPWL further reduced};
    \def\by{-5.2}
    \draw[block] (0,\by) rectangle (2.2,\by+1.2);
    \node[lbl] at (1.1,\by+0.9) {$b_1$};
    \draw[block] (2.8,\by+1.8) rectangle (5.0,\by+3.2);
    \node[lbl] at (3.9,\by+2.85) {$b_2'$};
    \node[pin] (p1b) at (1.85,\by+0.3) {};
    \node[pin] (p2b) at (3.1,\by+2.1) {};
    \draw[bbox] (1.85,\by+0.3) rectangle (3.1,\by+2.1);
    \draw[dim] (1.85,\by-0.4) -- node[below,lbl] {$\Delta x'$} (3.1,\by-0.4);
    \draw[dim] (5.2,\by+0.3) -- node[right,lbl] {$\Delta y'$} (5.2,\by+2.1);
    \draw[dotted,gray] (1.85,\by+0.3) -- (1.85,\by-0.4);
    \draw[dotted,gray] (3.1,\by+0.3) -- (3.1,\by-0.4);
    \draw[dotted,gray] (5.0,\by+0.3) -- (5.2,\by+0.3);
    \draw[dotted,gray] (5.0,\by+2.1) -- (5.2,\by+2.1);
    \node[rowlbl, anchor=east] at (1.0,\by+1.8) {rotation};
    \node[font=\small] at (2.5,\by-1.25) {(b) Pin-aware HPWL};
  \end{tikzpicture}
  \end{tabular}%
  }%
  \caption{Effect of block rotation on HPWL estimation.
    Crosses ($\times$) denote block centers and filled circles ($\bullet$) denote pin locations.
    (a)~Block-center HPWL remains unchanged after rotation, as it depends only on the
    rotation-invariant block center.
    (b)~Pin-aware HPWL captures the change in pin positions induced by rotation,
    enabling further wirelength reduction
    ($\Delta x' \!<\! \Delta x$, $\Delta y' \!<\! \Delta y$).}
  \label{fig:hpwl_comparison}
\end{figure}
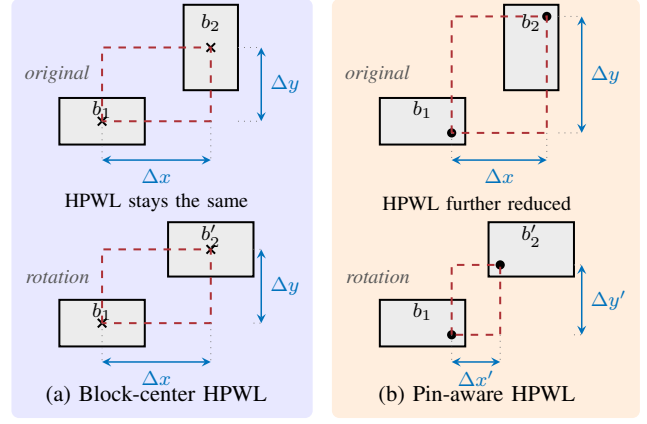

%% file: tables/table_notation.tex
\begin{table}[t!]
  \centering
  \footnotesize
  \tabcolsep=5pt
  \def\arraystretch{1.12}
  \caption{Notation.}
  \label{tab:notation}
  \resizebox{\linewidth}{!}{%
  \begin{tabular}{c | c}
    \toprule
    Symbol & Meaning \\
    \midrule
    $b_i \in B$;\; $w_i, h_i, p_i$ & block; its input width, height, power \\
    $n_j \in N$;\; $B_j$ & net; the blocks it connects \\
    $W, H$ & die outline dimensions (both layers) \\
    $(x_i, y_i)$;\; $z_i$ & bottom-left position; die assignment \\
    $o_i$;\; $f_i^h, f_i^v$ & orientation ($\mathrm{R0}$--$\mathrm{R270}$); flips \\
    $w'_i, h'_i$ & block dimensions after orientation \\
    $(u_{i,k}, v_{i,k})$;\; $(u'_{i,k}, v'_{i,k})$ & pin-$k$ center offset: input; transformed \\
    $(X_{i,k}, Y_{i,k})$ & global position of pin $k$ on block $b_i$ \\
    $T_r$;\; $T_f$ & rotation transform; flip transform \\
    $c_{i,q}$,\; $\mathcal{Q}_i$ & configuration binaries; feasible set \\
    $G$;\; $\eta$;\; $\mathcal{H}$ & thermal grid; power threshold; hot set \\
    $\rho$;\; $\mathcal{S}$;\; $\tau$ & thermal elasticity; its schedule; pruning threshold \\
    $\alpha$;\; $t_{\mathrm{peak}}$ & wirelength--thermal trade-off weight; field peak rise \\
    \bottomrule
  \end{tabular}}
\end{table}

%% file: doc/3-method.tex
\section{Methodology}
\label{sec:algo}
\subsection{Coarse Thermal Model and Thermal-First Placement}
\label{sec:stage1}
\label{sec:thermal_reward}
\begin{figure}[t!]
    \centering
    \includegraphics[width=0.98\linewidth]{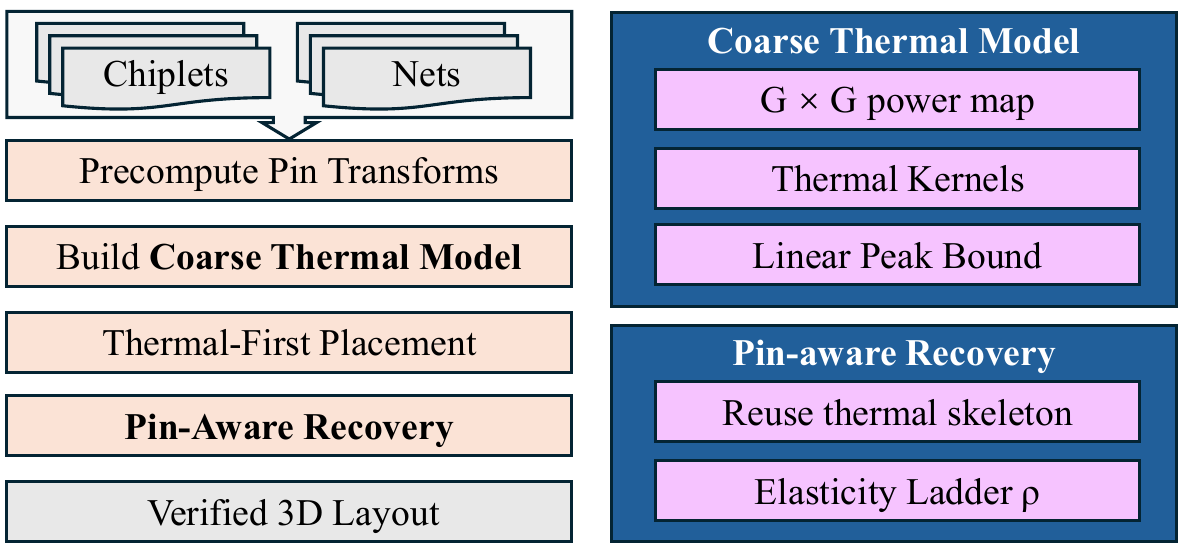}
    \caption{Overview of Chiplet3D: after an offline setup, Stage~1 fixes a cool
      thermal skeleton that Stage~2 refines for wirelength, with 3D-ICE
      reporting the final thermal distribution.}
    \label{fig:framework}
\end{figure}

A direct attack on \Cref{eq:objective} fails first at its temperature term: temperature has no closed form but comes out of a thermal simulator, far too expensive to call inside a search loop, so the simulator can validate layouts but never steer them.
As shown in \Cref{fig:framework}, Stage~1 sidesteps this by handing the solver a thermal model it can optimize directly.
At steady state heat conduction is linear, so the temperature rise on a die is just its power map convolved with a fixed point-spread kernel: each watt heats its own cell the most and nearby cells progressively less.
We embed a coarse form of this convolution, instead of a hand-crafted proxy such as simply spreading hot blocks apart, and use it to fix a cool thermal skeleton with orientation held fixed.
Partition each die into a $G\times G$ grid and let only the high-power blocks $\mathcal{H}=\{i : p_i \ge \eta\,p_{\max}\}$, the few that set the peak, enter the thermal model; the remaining blocks still obey the outline and non-overlap constraints but carry no thermal binaries.
Each $i\in\mathcal{H}$ takes a one-hot cell occupancy $g_{i,d,r,c}\in\{0,1\}$ that places it in exactly one cell of one die,
\begin{align}
  \sum_{d,r,c} g_{i,d,r,c}=1, \qquad \sum_{r,c} g_{i,1,r,c}=z_i, \label{eq:occ}
\end{align}
where the second equality ties the occupied die to the block's layer $z_i$ ($d{=}1$ is the sink-adjacent top die), and a big-M binds each block's center to its chosen cell.
The per-die power map is the occupancy weighted by power,
\begin{align}
  P_{d,r,c} = \sum_{i\in\mathcal{H}} p_i\, g_{i,d,r,c}, \label{eq:pmap}
\end{align}
and the resulting cell temperature rise is its convolution with the calibrated point-spread kernels,
\begin{align}
  \Theta_{d,r,c} = \sum_{d'}\sum_{r',c'} K^{\,d'\!\to d}_{\,r-r',\,c-c'}\, P_{d',r',c'}, \label{eq:conv}
\end{align}
where each $K^{d'\!\to d}$, one per source/target die pair, answers a concrete question: how much does one unit of power in a cell of die $d'$ heat every cell of die $d$? These responses are measured once, offline, from point-source simulations, so \Cref{eq:conv} captures heat spreading both laterally and vertically across the face-to-face stack.
A single variable then bounds every cell's rise from above,
\begin{align}
  t_{\mathrm{peak}} \;\ge\; \Theta_{d,r,c} \qquad \forall\, d,r,c, \label{eq:epi}
\end{align}
so minimizing $t_{\mathrm{peak}}$ presses down the hottest cell wherever it appears.
Stage~1 minimizes exactly this, namely \Cref{eq:objective} with $\alpha{=}0$ and orientation fixed, and returns the die and occupied cell of every hot block.

What \Cref{eq:conv} buys is an in-solver thermal field: it is linear in the occupancy binaries, so the MILP optimizes temperature directly during the search rather than checking it afterwards, at the cost of one coarse $G\times G$ convolution instead of a full simulation per candidate.

Coarseness costs resolution, not localization: on every case the surrogate pins the golden hottest cell exactly or to an adjacent cell (\Cref{tab:fidelity}), which is all the $\arg\min$ needs, since the deployed layout is always re-validated by the golden simulation in the end.
This completes Stage~1: every hot block holds a die and a cell, and this thermal skeleton is what Stage~2 preserves while it recovers wirelength.
\subsection{Pin-Aware Recovery with Dynamic Thermal Elasticity}
\label{sec:stage2}
\label{sec:pin_hpwl}
\label{sec:elastic}
\minisection{Pin geometry as a one-hot choice}
Stage~2 inherits the frozen skeleton and spends its remaining freedom on wirelength.
The obstacle is that the pin-aware HPWL of \Cref{eq:hpwl_pin} is nonlinear: rotating or flipping a block moves every pin (\Cref{eq:pin_global_prelim}), and a center-based proxy cannot even tell the orientations apart (\Cref{fig:hpwl_comparison}).
We linearize it by precomputing, for each block, the pin positions under all of its transforms before the solve, then letting the solver pick one.
Of the $16$ nominal rotation/flip combinations only $8$ are geometrically distinct, and small blocks, whose largest side falls below a threshold $\tau$, keep just the $4$ footprint-preserving mirrors, since a mirror relocates the whole pin bank without changing the block's footprint, whereas a dimension-swapping rotation only complicates non-overlap.
With one binary $c_{i,q}\in\{0,1\}$ per kept transform and $\sum_{q} c_{i,q}=1$, the pin offset is simply the selected one,
\begin{align}
  \delta_x^{i,k} = \sum_{q} c_{i,q}\, \delta_x^{i,k,q}, \qquad
  \delta_y^{i,k} = \sum_{q} c_{i,q}\, \delta_y^{i,k,q}, \label{eq:pin_select}
\end{align}
which stays linear because exactly one $c_{i,q}$ is active in \Cref{eq:pin_select}.
The same binaries set the block's orientation $o_i^{(r)}$ (one indicator per rotation $r$) and its flips,
\begin{align}
  o_i^{(r)} = \!\!\sum_{q:\,o(q)=r}\!\! c_{i,q}, \quad
  f_i^h = \!\!\sum_{q:\,f^h(q)=1}\!\! c_{i,q}, \quad
  f_i^v = \!\!\sum_{q:\,f^v(q)=1}\!\! c_{i,q}, \label{eq:conf_link}
\end{align}
so the placement constraints and the wirelength always see one consistent transform (\Cref{eq:conf_link}).

\minisection{Elasticity ladder}
Stage~2 minimizes the pin-aware HPWL over these transforms together with block positions, while holding the Stage-1 thermal field.
Writing $(\hat{x}_i,\hat{y}_i)$ for block $i$'s Stage-1 center, a single \emph{thermal-elasticity} coefficient $\rho\in[0,1]$ caps how far each hot block may move,
\begin{align}
  |x_i-\hat{x}_i|\le \tfrac{\rho}{2}\tfrac{W}{G}, \quad
  |y_i-\hat{y}_i|\le \tfrac{\rho}{2}\tfrac{H}{G} \quad \forall\, i\in\mathcal{H}, \label{eq:elastic}
\end{align}
with thermally negligible blocks left free. At $\rho{=}0$, \Cref{eq:elastic} freezes positions and only in-place transforms remain; at $\rho{=}1$ a hot block may cross a cell boundary, so every $\rho$ is re-verified rather than trusted.
We descend the elasticity schedule $\mathcal{S}=\{1,0.75,0.5,0.25,0.1,0\}$, reuse the single Stage-1 solve as a warm start, and keep the shortest layout whose re-estimated bottom-die peak passes the thermal gate; for each $\rho$ we also retain the rotation-only solution.
A larger $\rho$ trades thermal slack for shorter wires: thermally slack cases verify at $\rho{=}1$, while the tightest ones collapse toward $\rho{=}0$, where only the rotation-only candidate survives the gate.
Because Stage-1 is solved once and each rung is only a lightweight, warm-started re-solve plus one thermal-model check, the whole ladder stays well inside the Stage-2 budget (\Cref{sec:param_study}).

\subsection{End-to-End Flow and Efficient MILP Solving}
\label{sec:flow}
\label{sec:pruning}
Between them, the two stages recast every term of \Cref{eq:objective} into MILP form: the heat equation into a coarse linear convolution (\Cref{sec:stage1}), the pin-aware wirelength into a linear function of a precomputed configuration choice (\Cref{sec:stage2}), and non-overlap into the big-M constraints below.
Solving all thermal-occupancy and configuration binaries in one shot would still be heavy, which is exactly why the work is staged.
\Cref{alg:flow} assembles the two stages, each line tied to the equation it realizes.
The offline setup precomputes the per-transform pin offsets (line~\ref{line:pins}, \Cref{eq:pin_chain}) and calibrates the kernels (line~\ref{line:kernel}, \Cref{eq:conv}).
Stage~1 then selects the hot set (line~\ref{line:hot}), assigns each hot block a die and cell (line~\ref{line:occ}, \Cref{eq:occ}), and minimizes the coarse peak (line~\ref{line:stage1}, \Cref{eq:pmap,eq:conv,eq:epi}) to freeze the thermal skeleton (line~\ref{line:freeze}).
Stage~2 walks the elasticity ladder $\mathcal{S}$ (line~\ref{line:loop}): each rung picks one transform per block (line~\ref{line:cfg}, \Cref{eq:pin_select,eq:conf_link}), re-optimizes the pin-aware HPWL inside the $\rho$-box (line~\ref{line:reopt}, \Cref{eq:elastic}), and re-checks the layout's peak with the thermal model (line~\ref{line:verify}).
Throughout the flow, every temperature used to steer the search or select a layout comes from this coarse thermal model; the golden simulator enters only at the very end, where it independently validates and reports the final layouts in \Cref{sec:result}.

Read against the three limitations of \Cref{sec:intro}, the flow retires each by construction: the embedded field (line~\ref{line:stage1}) resolves hotspot geometry during the search itself~(3), and the per-transform pin offsets (line~\ref{line:cfg}) let the solver optimize true pin-to-pin wirelength~(1) over every geometrically distinct rotation and flip~(2).
The $600\,\si{\second}$ budget is split evenly between the two stages; the split is not sensitive, as Stage-1 either converges or returns a strong incumbent while Stage-2 finishes in seconds (\Cref{sec:param_study}).

With both objective terms linear, \Cref{eq:objective} under the placement constraints \eqref{eq:outline}--\eqref{eq:nonoverlap} is a standard MILP, and two choices keep its branch-and-bound search within budget.
First, non-overlap (\Cref{eq:nonoverlap}) demands that one of four relations hold: block $i$ lies to the left of, right of, below, or above block $j$. A standard big-M encoding with four edge-selection binaries $\sigma^{ij}_e\in\{0,1\}$ per same-die pair makes this choice linear,
\begin{equation}
\begin{aligned}
  x_i + w'_i &\le x_j + M\,\bar\sigma^{ij}_1, &
  y_i + h'_i &\le y_j + M\,\bar\sigma^{ij}_3, \\
  x_j + w'_j &\le x_i + M\,\bar\sigma^{ij}_2, &
  y_j + h'_j &\le y_i + M\,\bar\sigma^{ij}_4, \\
  \textstyle\sum_{e=1}^{4}\sigma^{ij}_e &\ge \gamma_{ij}, &
  M &= W+H,
\end{aligned}
\label{eq:bigM}
\end{equation}
where $\bar\sigma^{ij}_e=1-\sigma^{ij}_e$ and the same-die indicator $\gamma_{ij}\in\{0,1\}$ replaces the nonlinear test $z_i{=}z_j$: two blocks on the same die must keep at least one separating edge, while two blocks on different dies are left free to overlap in plan view, that is, to stack vertically, which is precisely the coupling the thermal field reasons about.
The constant $M{=}W{+}H$ is the smallest value the die outline allows, which keeps the solver's bounds tight and its pruning fast.

\input{doc/algs/two_stage_flow}
Second, removing redundant rotations/flips and pruning small blocks in \Cref{sec:stage2} cut the configuration binaries from a nominal $16n$ toward $4n$. We do not need to enforce a rigid limit on how power is split between the two dies. Instead, the cross-die thermal kernels in \Cref{eq:conv} naturally penalize stacking hot blocks vertically, while the self-kernels favor placing them on the top die (which is closer to the heat sink). As a result, the optimization objective itself automatically determines the best die assignment: the solver simply puts the entire hot set on the top die in five cases, and splits it across both dies in four cases, always choosing whichever setup stays cooler, a flexibility that a hard power-splitting constraint would otherwise block.

%% file: doc/algs/two_stage_flow.tex
\providecommand{\eqnum}[1]{(\ref{#1})}
\providecommand{\eqrange}[2]{(\ref{#1}--\ref{#2})}
\begin{algorithm}[t!]
\caption{The Chiplet3D two-stage flow}
\label{alg:flow}
\begin{algorithmic}[1]
\Require netlist $\mathcal{F}$; Chiplets; schedule $\mathcal{S}$
\Ensure a legal two-die layout that is short and cool
\Statex \(\triangleright\)~\emph{One-time setup}
\State precompute pin offsets $\delta^{i,k,q}$ \Comment{see \eqnum{eq:pin_chain}}\label{line:pins}
\State calibrate kernels $K^{d'\!\to d}$  \Comment{see \eqnum{eq:conv}}\label{line:kernel}
\Statex \(\triangleright\)~\emph{Stage 1: arrange the hot blocks to stay cool}
\State $\mathcal{H}\gets\{\,i : p_i\ge\eta\,p_{\max}\,\}$ \Comment{hot blocks}\label{line:hot}
\State assign each $i\in\mathcal{H}$ a die and cell \Comment{see \eqnum{eq:occ}}\label{line:occ}
\State $(z^\star,g^\star)\gets\arg\min\,t_{\mathrm{peak}}$ \Comment{see \eqrange{eq:pmap}{eq:epi}}\label{line:stage1}
\State freeze $(z^\star,g^\star)$ as the \emph{skeleton}\label{line:freeze}
\Statex \(\triangleright\)~\emph{Stage 2: shorten the wires without worse thermal}
\For{$\rho \in \mathcal{S}$, descending}\label{line:loop}
  \State pick one transform per block \Comment{see \eqrange{eq:pin_select}{eq:conf_link}}\label{line:cfg}
  \State $L_\rho\gets\arg\min\,\mathrm{HPWL}_{\mathrm{pin}}$ \Comment{see \eqnum{eq:elastic}}\label{line:reopt}
  \State $\widehat{t}\gets$ Thermal model; \label{line:verify}
\EndFor
\end{algorithmic}
\end{algorithm}

%% file: doc/4-exp.tex
\section{Experimental Results}
\label{sec:result}
\subsection{Experimental Setup}
\label{sec:exp_setup}
\noindent All experiments were conducted on a Linux workstation running Ubuntu 22.04 LTS, equipped with dual Intel Xeon Gold 6426Y processors and 256\,GB RAM, using
Python 3.12.2,
and Gurobi 11.0.3 as the MILP solver.
We evaluated on the benchmark suite from~\cite{2024ICCAD-ATPlace2.5D}.
Although originally designed for 2.5D chiplet placement on an interposer, we adapted each case to a 3D floorplanning problem, following the formulation in~\Cref{sec:problem_def}. Specifically, chiplets are mapped to two vertically stacked active dies.
The final layout of every method is evaluated by the golden thermal simulator 3D-ICE~\cite{zhu20253d}, all under the identical simulation setup listed in \Cref{tab:thermal_setup}.
\input{tables/table_thermal_setup}

\minisection{Metrics}
\Cref{tab:t600} reports three lower-is-better metrics.
(1) Wirelength is the pin-aware HPWL of \Cref{eq:hpwl_pin}, outline-normalized as $\mathrm{HPWL}_{\mathrm{norm}}=\mathrm{HPWL}/(W+H)$ so that cases of different die size are comparable;
(2) Peak temperature is the hottest point that 3D-ICE~\cite{zhu20253d} reports on either die; every method's layout is simulated under the same setup of \Cref{tab:thermal_setup}.
(3) Thermal uniformity measures how evenly heat is spread: we use the coefficient of variation $\mathrm{CoV}=\sigma/(\mu-T_{\mathrm{amb}})$ of the temperature rise over the active cells of both dies, a standard statistical measure of dispersion~\cite{abdi2010coefficient}.

\subsection{Comparison with State-of-the-Art Methods}
\label{sec:main_results}
\input{tables/table_T600}

We compare our method against two baselines: Lin~\cite{2021TVLSI_t3dflp}, a wirelength-first 3D floorplanner that treats temperature as a secondary objective, and Guan~\cite{2023TVLSI_t3dflp}, a thermal-aware 3D floorplanner. Both baselines run on the same adapted benchmarks, the same machine, and the same $600\,\si{\second}$ budget, and each is reported as the median over 10 random seeds. \Cref{tab:t600} reports the experimental results in terms of three metrics: peak temperature, thermal uniformity, and wirelength.

\minisection{Peak Temperature}
The peak always occurs on the bottom die, as it is farthest from the top-mounted heat sink. 
Steered by the thermal gate (\Cref{sec:elastic}), our Chiplet3D achieves strictly lower temperatures in this bottleneck die in all 10 cases compared to Guan’s method.

\begin{figure}[t!]
  \centering
  \includegraphics[width=0.95\linewidth]{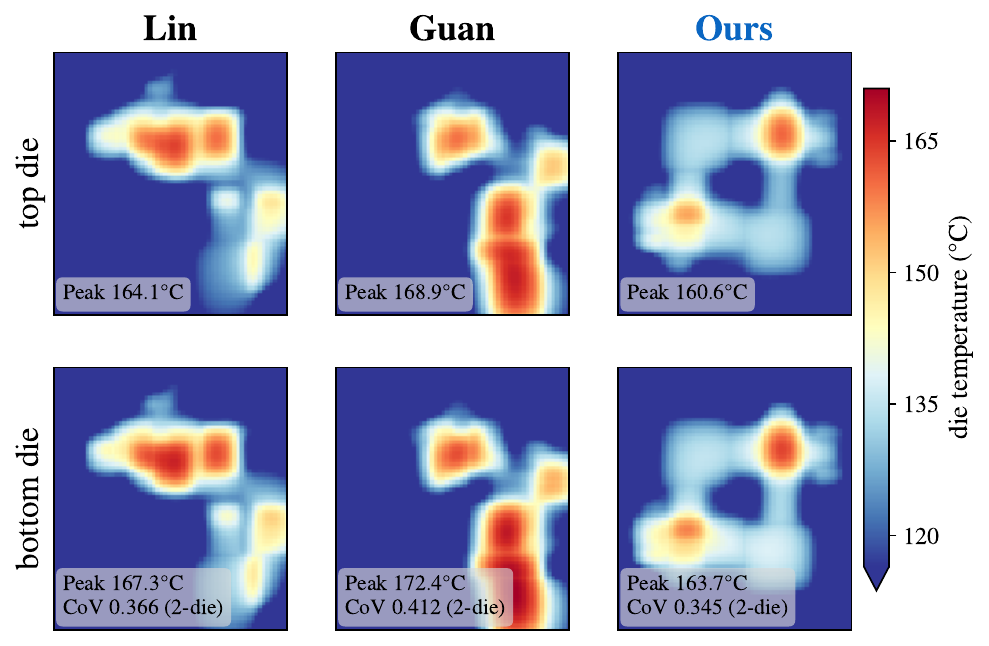}
  \caption{3D-ICE thermal maps of both dies for Case~7. Each panel reports the peak and the two-die CoV. Baseline values are 10-seed medians on a representative median-seed layout.}
  \label{fig:thermal_cmp}
\end{figure}

Against both baselines, ours achieves the lowest bottom-die peak temperature in 8 of 10 cases, and the two exceptions (Cases~2 and~10) are within $0.3\,\si{\celsius}$ of the best; on average, Lin and Guan are $1.13\times$ and $1.04\times$ hotter, respectively.
The margin widens on the hotspot-dominated cases: on Case~5, ours holds the bottom die to $162.7\,\si{\celsius}$ against Guan's $193.2$ and Lin's $208.6$, a $30$--$46\,\si{\celsius}$ reduction.
\Cref{fig:thermal_cmp} illustrates this for Case~7. Under identical budgets and color scales, both baselines (at their median seed) exhibit concentrated red hotspots, whereas our method keeps both dies cooler ($163.7$ vs.\ $167.3$/$172.4\,\si{\celsius}$) and produces flatter thermal maps.
Even the most challenging, thermally diffuse Case~6, our approach reduces the bottleneck die peak ($153.7$ vs.\ $156.0\,\si{\celsius}$).
This improvement comes from the coarse convolution field (\Cref{sec:thermal_reward}), which minimizes a faithful linear surrogate of the 3D-ICE peak, as verified in \Cref{sec:components}.

\minisection{Thermal Uniformity}
Beyond the peak, our method also spreads heat more evenly, attaining the lowest CoV in 9 of 10 cases, where Lin and Guan run $1.26\times$ and $1.14\times$ higher; the single exception, Case~8, goes to a Lin layout that runs about $10\,\si{\celsius}$ hotter, where the higher mean rise itself deflates CoV. This is a direct consequence of the coarse convolution field (\Cref{eq:conv}): minimizing the peak rise $t_{\mathrm{peak}}$ rewards moving power off the hottest cells into cooler ones, which flattens the whole on-die gradient rather than merely clipping its maximum, as the flatter maps for Case~7 in \Cref{fig:thermal_cmp} confirm.

\minisection{Wirelength}
Our method also achieves the shortest average pin-aware HPWL: Lin and Guan are $1.76\times$ and $1.64\times$ longer, respectively. Thus the thermal gains come without any wirelength penalty; the sole exception, Case~6, is a convergence limit rather than a model flaw, as \Cref{sec:components} shows.
Two factors contribute to this result. First, pin-aware modeling (\Cref{sec:pin_hpwl}) allows the solver to rotate and flip blocks, effectively pulling connected pins together (\Cref{fig:hpwl_comparison}) in the optimization procedure. Second, one global MILP floorplan all blocks and selects all transforms together, avoiding the quality loss of optimizing them piece by piece.

\subsection{Component Validation}
\label{sec:components}
We now evaluate in isolation the two key components underlying our method, pin-aware wirelength modeling and the coarse convolution thermal field.

\begin{figure}[t!]
  \centering
  \includegraphics[width=0.818\linewidth]{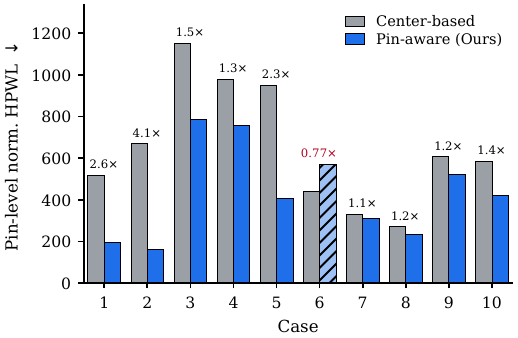}
  \caption{Pin-aware vs.~center-based HPWL under the same 600\,s runtime budget, both
    evaluated at true pin positions. The hatched bar for Case~6 indicates a convergence limit.}
  \label{fig:pin_ablation}
\end{figure}

\minisection{Effect of Pin-aware Wirelength Modeling}
We compare two wirelength models, pin-aware (using true pin coordinates) and center-based. All other components, including the thermal field, constraints, and orientation variables, remain unchanged. To ensure a fair comparison, both models are evaluated at the true pin positions.
As shown in \Cref{fig:pin_ablation}, the pin-aware model reduces wirelength in 9 of 10 cases. Its geometric-mean improvement over the center‑based model is $1.54\times$, and it achieves up to $4.1\times$ reduction on the small, orientation-dominated Cases~1--2. The pin‑aware wirelength model alone addresses the first two weaknesses identified in \Cref{sec:intro}, making the full transform space a useful decision variable.
For Case~4, the center-based run left a residual spacing violation, so its reported wirelength is a lower bound; fixing the violation would only lengthen it. The cost of pin‑aware modeling, specifically more configuration binaries and a looser relaxation, appears only on the thermally diffuse Case~6, whose nearly flat thermal field gives the solver little guidance, so the larger pin-aware model does not converge within 600\,s.
Consequently, Case~6 represents a convergence limit rather than a flaw in the pin-aware model.

\input{tables/table_fidelity}
\minisection{Thermal-field Fidelity}
The thermal improvements achieved by our method rely on the coarse convolution field accurately approximating the golden 3D‑ICE solution. \Cref{tab:fidelity} validates this fidelity on each case's most thermally stressed candidate layout, i.e., one that still contains a real hotspot. These are deliberately hot stress-test layouts, warmer than the cool result our method finally deploys (\Cref{tab:t600}), so the table probes surrogate accuracy rather than end-to-end quality; the surrogate field is compared to the golden 3D-ICE field on the same coarse grid.
The surrogate's peak cell matches the golden peak cell in 6 of 10 cases ($\Delta_{\mathrm{pk}}{=}0$) and an immediately adjacent cell in the other 4 ($\Delta_{\mathrm{pk}}{=}1$: edge-adjacent; $1.4{\approx}\sqrt{2}$: diagonal), never farther. The cell-wise field correlation (\emph{F}\,$r$) averages $0.72$ across both dies' coarse-grid cells.
This faithful tracking is exactly what closes the third limitation of prior methods listed in \Cref{sec:intro}: their simplified in-optimizer proxies see the true hotspot geometry only at post-placement validation, whereas our embedded field already follows the golden solution during the search.
Because the coarse grid lumps each block’s power at its center, the match is per cell rather than exact. This is why every reported temperature is a full 3D-ICE solve of the final layout.
\subsection{Parameter Studies and Trade-offs}
\label{sec:param_study}
\input{tables/table_budget_split}
\begin{figure}[t!]
  \centering
  \subfloat[Case~5]{\includegraphics[width=0.49\linewidth]{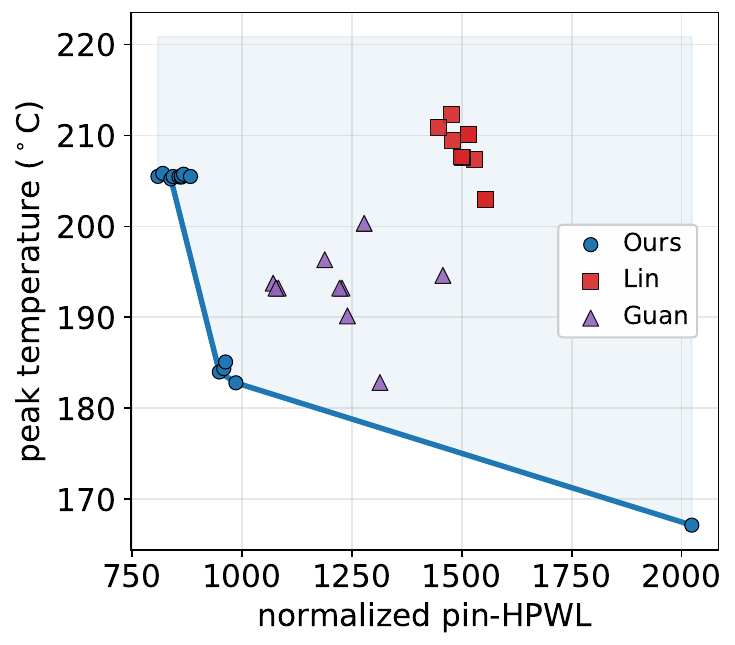}\label{fig:pareto_c5}}
  \hfill
  \subfloat[Case~4]{\includegraphics[width=0.49\linewidth]{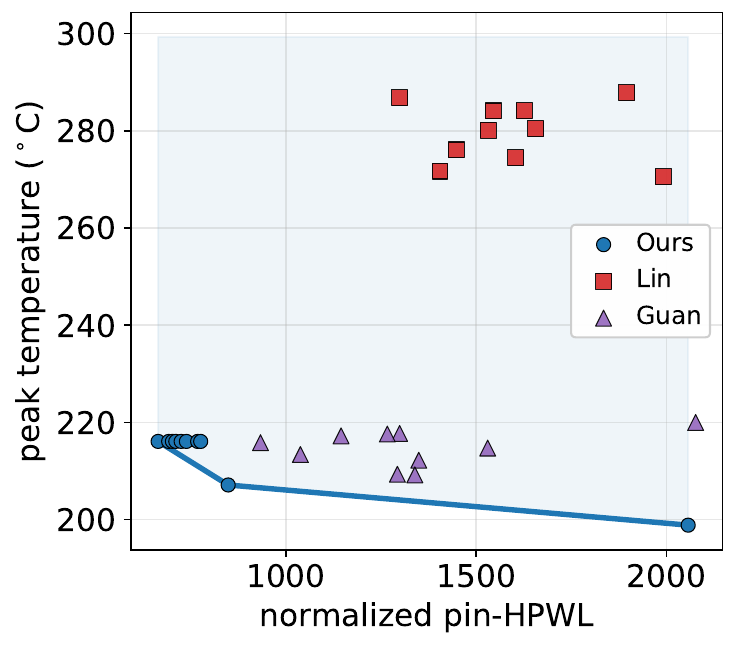}\label{fig:pareto_c4}}
  \caption{Per-case Pareto frontiers of wirelength--thermal trade-off. The curves show 3D-ICE peak temperature vs.\ normalized
    pin‑aware HPWL (lower-left is better). Baseline results are shown as seed clouds, and the shaded region indicates where our method dominates.}
  \label{fig:pareto}
\end{figure}
\minisection{Sensitivity to Stage Time Budget}
\Cref{tab:budget_split} sweeps the Stage-1/Stage-2 split across nine points of the fixed $600\,\si{\second}$ budget, reporting each case's bottom-die peak and pin-aware HPWL relative to its own adopted $300/300$, as the mean and worst case over the ten benchmarks.
The mean deviation is essentially zero throughout: cases whose Stage-1 already converges are unaffected by how the budget is divided.
Quality drops only when Stage-1 is starved (the leftmost columns), where the wirelength inflates by up to $34.6\%$ and the peak by up to $6.7\,\si{\celsius}$ as the under-resourced first stage returns a worse thermal skeleton.
Once Stage-1 receives at least $250\,\si{\second}$ the result is flat, with the worst case within $+2.2\,\si{\celsius}$ and $+1.1\%$ of the adopted point, so the even $300/300$ split sits comfortably inside this stable window.
Stage-2 is never the bottleneck: it proves optimality within $153\,\si{\second}$, and shrinking $t_2$ costs only a single optimality certificate (Case~2, whose incumbent is already the proven optimum).

\minisection{Pareto frontier of Wirelength–Thermal Trade-off}
\Cref{fig:pareto} sweeps the trade-off weight $\alpha$ of \Cref{eq:objective}. The curve links operating points obtained by our method, forming a per-case frontier from a cool long extreme to a short hot one.
On two representative cases, our frontier reaches the lower-left corner and dominates every baseline run (each shown as a 10-seed cloud) on both axes simultaneously, demonstrating robustness to run-to-run variance. For Case~5 (\Cref{fig:pareto_c5}), the frontier spans from a cool point at $167\,\si{\celsius}$ to a short point at a normalized HPWL of $810$. For Case~4 (\Cref{fig:pareto_c4}), Lin's method overheats ($>\!270\,\si{\celsius}$) while ours stays $\le\!216\,\si{\celsius}$ at a much shorter wirelength.
The main-table entry (\Cref{tab:t600}) is the coolest point of this frontier refined by Stage-2 (\Cref{sec:elastic}); the single weight $\alpha$ thus acts as a controllable dial between cooler layouts and shorter wires.

%% file: tables/table_thermal_setup.tex
\begin{table}[t]
  \centering
  \footnotesize
  \setlength{\tabcolsep}{4.5pt}
  \def\arraystretch{1.12}
  \caption{Illustration of used thermal simulation parameters in 3D-ICE. Thermal conductivity $k$ in
    \si{\watt\per\meter\per\kelvin}; volumetric heat capacity $c_v$ in
    \si{\mega\joule\per\cubic\meter\per\kelvin}.}
  \label{tab:thermal_setup}
  \begin{tabular}{l c c c l}
    \toprule
    Layer & Thickness & $k$ & $c_v$ & Role \\
    \midrule
    Heat sink (Cu)      & $6.9\,\si{\milli\meter}$ & $400$   & $3.55$ & sink \\
    Heat spreader (Cu)  & $1\,\si{\milli\meter}$   & $400$   & $3.55$ & spreader \\
    TIM                 & $100\,\si{\micro\meter}$ & $4.0$   & $4.00$ & passive \\
    Top die (Si)        & $100\,\si{\micro\meter}$ & $100$   & $1.75$ & active (tier 2) \\
    F2F bond   & $3\,\si{\micro\meter}$   & $1.6$   & $2.32$ & passive \\
    Bottom die (Si)     & $100\,\si{\micro\meter}$ & $100$   & $1.75$ & active (tier 1) \\
    Substrate           & $100\,\si{\micro\meter}$ & $0.300$ & $1.06$ & passive \\
    \midrule
    \multicolumn{5}{l}{Solver: 3D-ICE~\cite{zhu20253d}, steady state; peak $=\max$ over both dies.} \\
    \multicolumn{5}{l}{Ambient $T_{\mathrm{amb}}=45\,\si{\celsius}$ ($318.15\,\si{\kelvin}$); initial $=$ ambient.} \\
    \multicolumn{5}{l}{Grid: uniform $500\times500\,\si{\micro\meter\squared}$ cells; power by area overlap.} \\
    \multicolumn{5}{l}{Sink footprint $2(W{+}H)$; top convection from the contest resistance.} \\
    \bottomrule
  \end{tabular}
\end{table}

%% file: tables/table_T600.tex
\begin{table*}[t!]
  \centering
  \footnotesize
  \tabcolsep=3pt
  \def\arraystretch{1.2}
  \caption{Performance comparison on ATPlace Benchmark~\cite{2024ICCAD-ATPlace2.5D} under a 600\,s runtime budget. \textbf{Bold} marks cases where \textbf{Ours} achieves the best result. Lower values are better for HPWL, Peak temperature and CoV metrics.}
  \label{tab:t600}
  \centerline{\resizebox{1\linewidth}{!}{%
  \begin{tabular}{l cccc cccc cccc}
    \toprule
    & \multicolumn{4}{c}{Lin~\cite{2021TVLSI_t3dflp}}
    & \multicolumn{4}{c}{Guan~\cite{2023TVLSI_t3dflp}}
    & \multicolumn{4}{c}{\textbf{Ours}} \\
    \cmidrule(lr){2-5} \cmidrule(lr){6-9} \cmidrule(lr){10-13}
    Case & HPWL\,$\downarrow$ & Top Peak\,$\downarrow$ & Bot.\ Peak\,$\downarrow$ & CoV\,$\downarrow$
         & HPWL\,$\downarrow$ & Top Peak\,$\downarrow$ & Bot.\ Peak\,$\downarrow$ & CoV\,$\downarrow$
         & HPWL\,$\downarrow$ & Top Peak\,$\downarrow$ & Bot.\ Peak\,$\downarrow$ & CoV\,$\downarrow$ \\
    \midrule
    Case\,1 & \vr{548}{1.23} & \vr{162.3}{1.00} & \vr{164.9}{1.01} & \vr{0.209}{1.01} & \vr{617}{1.38} & \vr{161.8}{1.00} & \vr{164.3}{1.00} & \vr{0.225}{1.08} & \textbf{445} & 162.0 & \textbf{164.0} & \textbf{0.207} \\
    Case\,2 & \vr{307}{1.88} & \vr{105.3}{1.00} & \vr{106.0}{1.00} & \vr{0.384}{1.00} & \vr{307}{1.88} & \vr{105.3}{1.00} & \vr{106.9}{1.01} & \vr{0.388}{1.01} & \textbf{163} & 105.5 & 106.3 & \textbf{0.383} \\
    Case\,3 & \vr{2063}{2.69} & \vr{265.7}{1.36} & \vr{271.4}{1.39} & \vr{0.434}{2.37} & \vr{1622}{2.11} & \vr{199.3}{1.02} & \vr{199.9}{1.02} & \vr{0.194}{1.06} & \textbf{767} & \textbf{195.3} & \textbf{195.9} & \textbf{0.183} \\
    Case\,4 & \vr{1574}{2.43} & \vr{274.6}{1.34} & \vr{280.2}{1.37} & \vr{0.489}{1.29} & \vr{1296}{2.00} & \vr{209.6}{1.03} & \vr{215.2}{1.05} & \vr{0.406}{1.07} & \textbf{648} & \textbf{204.4} & \textbf{205.1} & \textbf{0.378} \\
    Case\,5 & \vr{1509}{2.16} & \vr{206.0}{1.27} & \vr{208.6}{1.28} & \vr{0.359}{2.29} & \vr{1225}{1.75} & \vr{190.7}{1.18} & \vr{193.2}{1.19} & \vr{0.279}{1.78} & \textbf{698} & \textbf{162.0} & \textbf{162.7} & \textbf{0.157} \\
    Case\,6 & \vr{703}{0.35} & \vr{186.4}{1.22} & \vr{190.1}{1.24} & \vr{0.302}{1.19} & \vr{661}{0.33} & \vr{152.4}{0.99} & \vr{156.0}{1.01} & \vr{0.257}{1.02} & 2033 & 153.3 & \textbf{153.7} & \textbf{0.253} \\
    Case\,7 & \vr{525}{1.99} & \vr{164.1}{1.02} & \vr{167.3}{1.02} & \vr{0.366}{1.06} & \vr{511}{1.94} & \vr{168.9}{1.05} & \vr{172.4}{1.05} & \vr{0.412}{1.20} & \textbf{264} & \textbf{160.6} & \textbf{163.7} & \textbf{0.345} \\
    Case\,8 & \vr{464}{2.31} & \vr{149.5}{1.06} & \vr{152.2}{1.07} & \vr{0.214}{0.80} & \vr{358}{1.78} & \vr{142.9}{1.02} & \vr{146.4}{1.03} & \vr{0.279}{1.04} & \textbf{201} & \textbf{140.5} & \textbf{141.9} & 0.268 \\
    Case\,9 & \vr{1155}{2.75} & \vr{223.5}{1.02} & \vr{226.8}{1.02} & \vr{0.249}{1.18} & \vr{1182}{2.82} & \vr{220.4}{1.01} & \vr{223.9}{1.01} & \vr{0.217}{1.03} & \textbf{420} & \textbf{218.4} & \textbf{221.6} & \textbf{0.211} \\
    Case\,10 & \vr{881}{2.03} & \vr{208.8}{1.00} & \vr{212.1}{1.00} & \vr{0.318}{1.17} & \vr{973}{2.24} & \vr{212.0}{1.02} & \vr{215.6}{1.02} & \vr{0.337}{1.24} & \textbf{434} & \textbf{208.8} & 212.2 & \textbf{0.272} \\
    \midrule
    Ratio & $1.76\times$ & $1.12\times$ & $1.13\times$ & $1.26\times$ & $1.64\times$ & $1.03\times$ & $1.04\times$ & $1.14\times$ & {\boldmath$1.00\times$} & {\boldmath$1.00\times$} & {\boldmath$1.00\times$} & {\boldmath$1.00\times$} \\
    \bottomrule
  \end{tabular}}}
  {\footnotesize
  \par\vspace{3pt}
  \centerline{\parbox{0.95\linewidth}{\noindent\textit{*}~Top/Bot.\ Peak are the peak temperatures(\si{\celsius}) of the top and bottom dies, respectively, reported by 3D-ICE. HPWL and CoV are defined in \Cref{sec:exp_setup}.
  For each baseline, the reported value is the per-case ratio of its 10-seed median to our result. Our two-stage flow is deterministic and uses no random seed, so a single run per case is reported. The bottom row gives the geometric mean of
  these per-case ratios.}}
  }
\end{table*}

%% file: tables/table_fidelity.tex
\begin{table}[t!]
  \centering
  \footnotesize
  \tabcolsep=3.6pt
  \def\arraystretch{1.12}
  \caption{Fidelity of the coarse surrogate to the golden field on each case's hottest probe layout. $T_{\mathrm{pk}}$: golden bottom-die peak; $\Delta_{\mathrm{pk}}$: cell-grid distance from the surrogate's to the golden peak cell ($0$ = exact); F\,$r$: cell-wise field correlation.}
  \label{tab:fidelity}
  \resizebox{\linewidth}{!}{%
  \begin{tabular}{l cccc | l cccc}
    \toprule
    Case & $T_{\mathrm{pk}}$ & Range & $\Delta_{\mathrm{pk}}$ & \emph{F}\,$r$ & Case & $T_{\mathrm{pk}}$ & Range & $\Delta_{\mathrm{pk}}$ & \emph{F}\,$r$ \\
    \midrule
    C1 & 168 & 162--168 & 1.4 & $+0.70$ & C6  & 198 & 154--198 & 0 & $+0.97$ \\
    C2 & 106 & 105--106 & 0   & $+0.76$ & C7  & 172 & 146--172 & 0 & $+0.65$ \\
    C3 & 278 & 196--278 & 0   & $+0.75$ & C8  & 158 & 138--158 & 0 & $+0.69$ \\
    C4 & 203 & 196--203 & 1   & $+0.89$ & C9  & 227 & 200--227 & 0 & $+0.82$ \\
    C5 & 196 & 160--196 & 1.4 & $+0.51$ & C10 & 209 & 185--209 & 1 & $+0.43$ \\
    \bottomrule
  \end{tabular}}
\end{table}

%% file: tables/table_budget_split.tex
\begin{table}[t]
  \centering
  \footnotesize
  \def\arraystretch{1.40}
  \setlength{\tabcolsep}{3pt}
  \caption{Sensitivity to the Stage-1/Stage-2 time split showing deviations of the bottom-die peak
    ($\Delta T_{\mathrm{bot}}$, \si{\celsius}) and pin-aware HPWL ($\Delta$HPWL, \%) from the $300/300$ reference split versus the Stage-1 time $t_1$ ($t_2=600-t_1$).}
  \label{tab:budget_split}
  \resizebox{\linewidth}{!}{%
  \begin{tabular}{l ccccccccc}
    \toprule
    $t_1$ (s) & $60$ & $150$ & $200$ & $250$ & $\mathbf{300}$ & $350$ & $400$ & $450$ & $540$ \\
    \midrule
    $\Delta T_{\mathrm{bot}}$ mean & $+1.4$ & $+1.1$ & $+0.0$ & $+0.1$ & $\mathbf{0.0}$ & $+0.2$ & $+0.2$ & $+0.2$ & $+0.4$ \\
    $\Delta T_{\mathrm{bot}}$ max  & $+6.1$ & $+6.7$ & $+2.4$ & $+0.8$ & $\mathbf{0.0}$ & $+2.2$ & $+2.2$ & $+2.2$ & $+2.2$ \\
    $\Delta$HPWL mean             & $+10.9$ & $+1.4$ & $+0.2$ & $+0.0$ & $\mathbf{0.0}$ & $-0.1$ & $-0.1$ & $-0.1$ & $-0.1$ \\
    $\Delta$HPWL max              & $+34.6$ & $+9.3$ & $+2.2$ & $+0.1$ & $\mathbf{0.0}$ & $+0.0$ & $+0.0$ & $+0.0$ & $+1.1$ \\
    \bottomrule
  \end{tabular}}
\end{table}

%% file: doc/5-conclusion.tex
\section{Conclusion}
\label{sec:conclu}

This paper presented Chiplet3D, a pin- and thermal-aware MILP floorplanner for two-die 3-D chiplet integration. 
Chiplet3D models true pin locations under rotations and flips, embeds a compact convolution-based thermal field inside the solver, and uses a two-stage flow to recover pin-aware wirelength while preserving thermal quality. 
On the ICCAD'24 ATPlace benchmarks,
with every temperature validated by golden 3D-ICE simulation,
Chiplet3D achieves the lowest peak temperature on 8 of 10 cases, the most uniform thermal profile on 9 of 10 cases, and the shortest average wirelength. 
Compared with SOTA baselines, it reduces wirelength by 39\%--43\% on average and up to 62\%, while lowering peak temperature by up to 45.9$^\circ$C and thermal non-uniformity by up to 56\%. 
These results show that co-optimizing pin alignment and thermal fields yields a stronger wirelength and thermal Pareto frontier for 3D chiplet floorplanning.